\documentclass[12pt]{article}
\usepackage[english]{babel} 
\usepackage[utf8]{inputenc}
\usepackage[T1]{fontenc}

\usepackage{graphicx} 
\usepackage{amsmath}
\usepackage{amssymb} 
\usepackage{mathtools} 
\usepackage{bm}
\usepackage{tabularx}
\usepackage{booktabs}

\DeclarePairedDelimiter{\abs}{\lvert}{\rvert}
\newcommand{\vect}[1]{\mathbf{#1}}
\newcommand{\mvec}{\mathbf}
\newcommand{\scripts}[1]{\scriptscriptstyle{#1}}
\newcommand{\latt}[2]{ \mvec{#1}_{\mvec{#2}} }
\newcommand{\scrlatt}[2]{ \mvec{#1}_{\mvec{#2}} }
\newcommand{\lattb}[3]{ \mvec{ #1}_{\mvec{#2}}^{\scripts{( #3 )} } }
\newcommand{\scrlattb}[3]{ \mvec{ #1}_{\mvec{#2}}^{( #3 )} }
\newcommand{\lattf}[3]{\mvec{F}_{\mvec{#1},\scripts{(#2)}}^{\text{#3}}}
\newcommand{\lattser}[2]{\sum_{(\mvec{#1},#2)}}

\newcommand{\ord}[1]{\scriptscriptstyle{[#1]}}
\newcommand{\curl}{\text{\bf{curl}}}

\newcommand{\definedas}{\coloneqq}
\newcommand{\definedasright}{\eqqcolon}
\newcommand{\dif}{\mathrm{d}}
\DeclareMathOperator{\erfc}{erfc}
\DeclareMathOperator{\erf}{erf}
\title{Classical microscopic theory of polaritons in ionic crystals}

\author{ A.~Lerose\thanks{School of Physics, Universit\`a degli Studi di%
   Milano, Italy}%
 \and A.~Sanzeni\footnotemark[1]%
 \and A.~Carati\thanks{Dipartimento di Matematica, Via Saldini 50,
   Milano, I--20133, Italy, E--mail: \texttt{andrea.carati@unimi.it}.}%
 \and L. Galgani\footnotemark[2]%
}

\date{\today}
\begin{document}
\maketitle

\begin{abstract}
It is well known that the optical branches of the dispersion curves of
ionic crystals exhibit a polaritonic feature, i.e., 
a splitting about the electromagnetic
dispersion line $\omega=ck$. This phenomenon is considered to be due
to the  retardation of the electromagnetic forces among the ions. However,
the problem is  usually discussed  at a
phenomenological level, through the introduction of a macroscopic polarization
field, so that 
a microscopic treatment is apparently lacking. A microscopic first
principles deduction is given here, in a classical frame, for a
model in which 
the ions are dealt with as point charges. At a
qualitative level it is made apparent  that  retardation is indeed
responsible for the splitting.  
A quantitative comparison with the empirical data for LiF is also
given, showing a fairly good agreement over the whole Brillouin zone.
\end{abstract}


\newpage

\section{Introduction}

The existence of polariton  dispersion curves in ionic
crystals is of great physical relevance, inasmuch as it allows one to
explain the phenomenon of the dispersion of light in such
crystals. Polaritonic curves exist only if one takes into account the
retarded nature of the electromagnetic  forces among the ions in the crystal. 
Indeed, if retardation is neglected one obtains the familiar Coulomb model
in which the dispersion curves of the crystal lattice and those of light  are
completely different, whereas they coincide  along the polaritonic
lines. So retardation is the essential necessary ingredient
in order to explain the optics of crystals,  and it will be shown here
that it is the retarded interaction with the \emph{far} ions that is
actually responsible for the  splitting.

Now,  a splitting of the  dispersion curves of an ionic crystal  
about the  electromagnetic line  $\omega=ck$, was
apparently first predicted   by Born and Huang (see
\cite{bornhuang}, pages 91 and  94) in the frame of a
phenomenological discussion of the problem in terms of a macroscopic 
polarization field. Further discussions were then
given  by Fano \cite{fano} and Hopfield   
\cite{hopfield}, still in terms of macroscopic polarization densities.  
The splitting
was finally  observed in the years  60s, first in semiconductors and
later in LiF and in other ionic crystals.
In conclusion, it is now commonly  assumed  (see for example 
\cite{polaritons}, sections 7.2 and
7.3) that the  phenomenon should be
understood as due to retardation. However, retardation is usually
 introduced through the
phenomenological Maxwell equations which involve, in addition to
the microscopic field $\mvec{E}$, also the macroscopic field
$\mvec{D}$. A  microscopic deduction 
 is thus apparently lacking.

In the present paper such a microscopic deduction is given,  in a classical
frame.
We consider a model in which the ions are dealt with as point particles,
 internal degrees of freedom being neglected.  The existence of
 molecular repulsive forces balancing the Coulomb ones, and thus 
allowing for the existence of a lattice, 
 is assumed at a phenomenological level.  
The Newton equations of motion for the displacements of the ions, with 
retardation of the electromagnetic forces taken into account, are written down 
in the linear approximation. At the end the model involves as free 
parameters  the  constants entering the repulsive forces, and
 the effective charge of each ion, in addition to 
 the geometric parameters of the
 lattice and to the  ions' masses.
 The  normal modes
are  numerically determined for crystals with a rock salt
structure, 
through a suitable procedure, which is required in order 
to take the effect of retardation into account. The three free
parameters are determined by a comparison with the experimental data
of the dispersion curves of LiF, while the remaining parameters were
taken from the literature.  Finally, 
the electromagnetic field generated by the
motions of the ions is discussed.

The main results are the following:
\begin{enumerate}
\item In the dispersion curves of the lattice vibrations there appear  branches 
that are absent in the purely \emph{``mechanical or instantaneous'' model}
(in which retardation  is neglected), and correspond to the 
previously mentioned  polaritonic splitting. The splitting
turns out to be actually  due to the retarded
interaction with the \emph{far} ions.  

\item The agreement  between the theoretically computed 
dispersion curves and the experimental ones available in the
literature for the case of LiF (see
figures~\ref{fig:1} and ~\ref{fig:2}) is, in our opinion, fairly good for
all values of $\mvec{k}$ in the whole Brillouin zone. This is obtained
with no need of
introducing a phenomenological value for the static dielectric 
constant $\varepsilon$,  which is here deduced from the theory.

\item The electromagnetic field created by the motion of the
 ions can be decomposed into a microscopic part and
a macroscopic one. The former propagates at the vacuum speed $c$, while 
the latter propagates  according to the laws of
macroscopic optics, with a  phase velocity $\omega (\mvec{k})/k$. 

\item The dispersion relation  $\omega(\mvec{k})$  of the macroscopic 
electric field coincides with 
the vibrational one  of the  lattice.
\end{enumerate}

In Section~2 the microscopic model is described and the linearized equations of
motion are obtained. In Section~3
it is shown how, due to retardation,  the secular
equation presents a peculiar form,
 which is responsible for the occurrence of the splitting.
In Section~4 the dispersion curves explicitly computed
for rock salt lattices are reported, and the comparison with the
experimental data is performed. 
Some further general problems 
concerning the  microscopic deduction of optics are discussed in
Section~5. More  comments are given in a conclusive section.
Appendix \ref{app:WF} is devoted to recalling   the role played by 
the so called
Wheeler--Feynman identity\cite{wf} (which is a theorem in the present  model) 
in ensuring the stability of the lattice. Indeed, such an identity
 guarantees that the microscopic  dispersion  relations do not contain 
imaginary terms, if  the familiar 
\emph{radiation reaction force}\cite{jackson} acting on
each ion is taken into account.
Appendices \ref{app:ewald} and \ref{app:dynmatrix} contain  
details about some analytical computations discussed in the text.

\section{The model}
We consider a model in which the ions are described as point
charges, internal degrees of freedom being neglected. The ions 
 interact both through a 
phenomenological effective
potential, that accounts for the well-known short-range repulsive
quantum effects associated to the ``impenetrability'' 
of matter\footnote{A first-principle calculation of the repulsive short-range
  potential might  be attempted through standard quantum many-body
  methods.  However, here this is a minor issue, since our
  main concern is the correct treatment of the \emph{electromagnetic}
  interactions.}, and through
the forces due to the electromagnetic
field created by all the other ions. In addition, each ion is subject
to  the radiation reaction force. The latter force is included because,
although having a negligible magnitude, it  plays a
qualitatively relevant role  in making the theory  consistent, i.e.,
in ensuring the stability of the lattice.
So the Newton equation of each ion 
(with mass $m$ and  position vector
${\mvec x}$) has the form
$$
m\ddot{\mvec x}=\mvec F^{\mathrm{rep}} +\mvec F^{\mathrm{em}}
+\mvec F^{\mathrm{rr}}\ ,
$$
where $\mvec F^{\mathrm{rep}}$, $\mvec
F^{\mathrm{em}}$ and $\mvec F^{\mathrm{rr}}$ denote, respectively,  the
short-range repulsive force describing the  interaction with neighboring ions, 
 the electromagnetic force due to  all the other ions, and  the 
 radiation reaction force.

Actually the  model is studied  in its linearized
version. So, first of all we assume there exists an equilibrium configuration 
in which the ions sit on the lattice sites $\lattb{r}{h}{j}$ 
of the crystal under
consideration, determined by the repulsive forces and the Coulomb ones.
Here, as usual, 
${\mvec{ h}} \in \mathbb{Z}^3$ denotes
the cell, while $j=1,\dots,n$ denotes the ion's species. 

Thus, instead of 
 the actual positions $\lattb{x}{h}{j}$ of the ions, the 
 relevant quantities are the corresponding displacements
 $$
\lattb{u}{h}{j} =\lattb{x}{h}{j} - \lattb{r}{h}{j} \ .
$$  
So, the system of Newton equations for  the ions' motions takes the form
\begin{equation}
\label{eq:genericforce}
m^{\scripts{(j)}} \lattb{\ddot{u}}{h}{j} = \lattf{h}{j}{rep} +
\lattf{h}{j}{em} + \lattf{h}{j}{rr} \ , \quad (j=1,...,n)\ ,
\end{equation}
where $m^{\scripts{(j)}}$ is the mass of the ions of species $(j)$.

Then the equations  of motion  are linearized with respect to 
the displacements. 
So, to start with,  the magnetic field is completely neglected, and
 the Abraham-Lorentz-Dirac radiation reaction force is taken in its
nonrelativistic approximation, given by (see \cite{jackson})
\begin{equation} 
\label{eq:rr}
\tfrac{2}{3} \tfrac{ \left( q^{\scripts{(j)}} \right)^2 }{ c^3}
\lattb{\dddot{x}}{h}{j} =
\tfrac{2}{3} \tfrac{ \left( q^{\scripts{(j)}} \right)^2 }{ c^3}
\lattb{\dddot{u}}{h}{j} 
\end{equation}
where $q^{\scripts{(j)}}$ is the charge of the ions of species $(j)$. 
 
As far as the molecular repulsive forces are concerned, they may be assumed
to have a phenomenological simplified form corresponding to two-body central
potentials $\phi_{\scripts{(j,l)}}(r)$ (a priori different for each
pair of species). Then, the linearization procedure amounts to computing
the de\-rivatives of the total potential, evaluated at the equilibrium
configuration. Due to the short range of the repulsive
forces, such derivatives rapidly approach zero as the distance between particles
 becomes sufficiently large (roughly, for
distances  larger than the linear
dimension of the unit cell). In fact one  may assume that the
interactions occur only over the neighbors of the $(\mvec{h},j)$
ion. Denoting  by
\[ \mvec{r}_{\mvec{h} - \mvec{d}_j(s)}^{\scripts{(\tau_j(s))}},  
\qquad s=1,\dots,n_j \] 
the relative positions of the   $n_j$ neighbors of a
$j$-th species ion with respect to the ion $(\mvec{h},j)$,
it is easy to show that the linearized repulsive forces have the form
\begin{equation}
\label{eq:repulsiveforce}
\lattf{h}{j}{rep} = \sum_{s=1}^{n_j} \Bigl( \alpha_{(j),s}
\Pi_{(j),s}^{\parallel} + \beta_{(j),s} \Pi_{(j),s}^{\perp} \Bigr) \cdot
\bigl(\mvec{u}_{\mvec{h} - \mvec{d}_j(s)}^{\scripts{(\tau_j(s))}} -
\lattb{u}{h}{j}\bigl),
\end{equation}
where $\Pi_{(j),s}^{\parallel}$ and $\Pi_{(j),s}^{\perp}$ are
 the projection operators, respectively along the direction of the
neighbor separation vector
$\mvec{r}_{\mvec{d}_j(s)}^{\scripts{(j,\tau_j(s))}}$, and  onto the
plane normal to it. The parameters 
\[
\begin{split}
 \alpha_{(j),s} & \definedas
 \phi''_{\scripts{(j,\tau_j(s))}}\Bigl(\abs[\big]{\mvec{r}_{\mvec{d}_j(s)}^{\scripts{(j,\tau_j(s))}}}
 \Bigr) , \\
 \beta_{(j),s} & \definedas
 \frac{\phi'_{\scripts{(j,\tau_j(s))}}\Bigl(\abs[\big]{\mvec{r}_{\mvec{d}_j(s)}^{\scripts{(j,\tau_j(s))}}}
   \Bigr)}{\abs[\big]{\mvec{r}_{\mvec{d}_j(s)}^{\scripts{(j,\tau_j(s))}}}}
 ,
\end{split}
\] 
for $ j=1,\dots,n, \; s=1,\dots,n_j$, are characteristic of the
concrete crystal under consideration. Obviously, if a lattice presents a
non-trivial point symmetry group, some of the parameters are likely to
coincide.

Finally, we come to the linearization  of the electric  forces acting on
each ion and due to all the other ones  (the magnetic
forces having been neglected). The
 electromagnetic field created by the ions
 is taken in the dipole approximation, i.e., as the field obtained from
 the Maxwell equations when
the charge distribution and the current density of the sources are
linearized with respect to the displacements from their 
equilibrium positions. So the
linearized source corresponding to a certain ion, labeled by a cell 
index $\mvec{p}$ and a species index $(s)$, with actual motion
$\lattb{r}{p}{s} +\lattb{u}{p}{s}(t)$, is the superposition of a 
\emph{static source} $q^{\scripts{(s)}}$, 
of zero--th order 
in the displacement $\lattb{u}{p}{s}$,  and of a first--order \emph{dipole
 source}, with a dipole $\mvec{d}(t) = q \lattb{u}{p}{s}(t)$, 
both at the position $\lattb{r}{p}{s}$. 
The solution of the corresponding inhomogeneous 
Maxwell equations at the
spacetime point $(\mvec{x},t)$ is given
by the superposition of the respective fields: a static
spherically-symmetric \emph{Coulomb field}
\begin{equation}\label{eq:coulfield}
\mvec{E}^{\ord{0}}(\mvec{x},t) = q^{\scripts{(s)}} \, \nabla
\frac{1}{ \abs{\mvec{x}-\lattb{r}{p}{s}} }\ 
\end{equation}
 and a variable \emph{dipole field}
\begin{equation}
\label{eq:lindipef}
\mvec{E}^{\ord{1}}(\mvec{x},t) = q^{\scripts{(s)}} \, \curl\ \curl \, \frac{\lattb{u}{p}{s}
  \big(t - \frac{1}{c}
  \abs{\mvec{x}-\lattb{r}{p}{s}}\big)}{\abs{\mvec{x}-\lattb{r}{p}{s}}}\ ,
\end{equation}
which is the one involving retardation . 
We are using here the standard representation of the
dipole field that one finds in the classical works of Ewald
\cite{ewald}, Oseen \cite{oseen} and
Born \cite{optik} on the subject. For
a detailed derivation see for example \cite{bornwolf} 
or \cite{tikhonov}. Such 
electric terms will be referred to  as the ``Coulomb
term'' and the ``dipole term'' respectively. 

We now consider  the forces resulting from the action of such fields
upon a certain ion, identified by the cell index $\mvec{h}$ and the 
species index $(j)$, due to all the other ones, in the linear approximation 
(with respect to the displacements $\mvec{u}$'s). 
Its actual motion is similarly denoted $\lattb{r}{h}{j} + \lattb{u}{h}{j}(t) $.

The dipole field given by (\ref{eq:lindipef}), is already linear in $\lattb{u}{p}{s}$, 
hence the linear approximation of the resulting electric force 
amounts to evaluating it at $\mvec{x}=\lattb{r}{h}{j}$. 
Summing over all ions of the lattice, the force due to the dipoles is thus
\begin{equation}\label{eq:forzdip}
  \lattf{h}{j}{dip} =
  q^{\scripts{(j)}} \sideset{}{'}\sum_{(\mvec{p},s)} q^{\scripts{(s)}} \curl \,
  \curl \, \frac{\lattb{u}{p}{s} \Big(t - \frac{1}{c}
    \abs{\mvec{x}-\lattb{r}{p}{s}}\Big)}{\abs{\mvec{x}-\lattb{r}{p}{s}}}
  \Bigg\vert_{\mvec{x}=\lattb{r}{h}{j}} \ ,
\end{equation}
where the prime in the  sums denotes that the 
term $(\mvec{p},s)=(\mvec{h},j)$ is excluded.
For what concerns the force due to the static Coulomb field, one has
to take the force given by (\ref{eq:coulfield}), evaluate it at the point
$\lattb{r}{h}{j} + \lattb{u}{h}{j}(t) $, and expand it up to the first
order in $\lattb{u}{h}{j}(t) $.
The zeroth order term is balanced by the short range repulsive
contribution, because we are evaluating
the field at the
equilibrium configuration, while the linear term
can be written in a form which resembles that 
of the dipole term and highlights the electrostatic potential. 
In fact, the first order term is given by
$$
\lattb{u}{h}{j}(t)\cdot\nabla \Big(
\mvec{E}^{\ord{0}}(\lattb{r}{h}{j},t) \Big) \ ,
$$
where$\mvec{E}^{\ord{0}}$ is the gradient of the electrostatic
potential. Using the identity
\[
\big( \mvec{u}\cdot\nabla \big)\, \nabla \Phi = - \curl \,
\curl \ \big( \mvec{u} \, \Phi ( \mvec{x}) \big)\ ,
\]
which holds for any scalar field
$\Phi$, and taking for $\Phi$ the Coulomb potential,
the linearized electric force given by the Coulomb term  reads
\[
 - q^{\scripts{(j)}} q^{\scripts{(s)}} \, \curl \ \curl\
\frac{\lattb{u}{h}{j}(t)}{\abs{\mvec{x}-\mvec{r}}}
\bigg\vert_{\mvec{x}=\lattb{r}{h}{j}} \ .
\]
So,  summing over all ions of the lattice one gets
\begin{equation}
\label{eq:coulombseries}
 \lattf{h}{j}{coul} = - q^{\scripts{(j)}} \sideset{}{'}\sum_{(\mvec{p},s)}
 q^{\scripts{(s)}} \curl \, \curl \, \frac{ \lattb{u}{h}{j}(t) }{
   \abs[\big]{\mvec{x}-\lattb{r}{p}{s}}}
 \Bigg\vert_{\mvec{x}=\lattb{r}{h}{j}} ; \\ 
\end{equation}
where the prime in the  sum denotes that the 
term $(\mvec{p},s)=(\mvec{h},j)$ is excluded.

In conclusion, the linearized equations of motion for the
charges of our ionic lattice read
\begin{equation}
\label{eq:lineareq}
m^{\scripts{(j)}} \lattb{\ddot{u}}{h}{j}= \lattf{h}{j}{rep} +
\lattf{h}{j}{coul} + \lattf{h}{j}{dip} + \lattf{h}{j}{rr}\ ,
\end{equation}
with
\begin{align*}
\lattf{h}{j}{rep} &= \sum_{s=1}^{n_j} \Bigl( \alpha_{(j)s}
\Pi_{(j)s}^{\parallel} + \beta_{(j)s} \Pi_{(j)s}^{\perp} \Bigr) \cdot
\bigl(\mvec{u}_{\scripts{\mvec{h} -
    \mvec{d}_j(s)}}^{\scripts{\tau_j(s)}} - \lattb{u}{h}{j}\bigl)\ ; \\
\label{eq:coulombseries}
 \lattf{h}{j}{coul} &= - q^{\scripts{(j)}} \sideset{}{'}\sum_{(\mvec{p},s)}
 q^{\scripts{(s)}} \curl \, \curl \, \frac{ \lattb{u}{h}{j}(t) }{
   \abs[\big]{\mvec{x}-\lattb{r}{p}{s}}}
 \Bigg\vert_{\mvec{x}=\lattb{r}{h}{j}} ; \\ 
\lattf{h}{j}{dip} &=
 q^{\scripts{(j)}} \sideset{}{'}\sum_{(\mvec{p},s)} q^{\scripts{(s)}} \curl \,
 \curl \, \frac{\lattb{u}{p}{s} \Big(t - \frac{1}{c}
   \abs{\mvec{x}-\lattb{r}{p}{s}}\Big)}{\abs{\mvec{x}-\lattb{r}{p}{s}}}
 \Bigg\vert_{\mvec{x}=\lattb{r}{h}{j}} ; \\ 
\lattf{h}{j}{rr} &=
 \frac{2}{3} \frac{ \left( q^{\scripts{(j)}} \right)^2 }{ c^3}
 \lattb{\dddot{u}}{h}{j} \ ,
\end{align*}
where the prime in the  sums denotes that the 
term $(\mvec{p},s)=(\mvec{h},j)$ is excluded.
Equations (\ref{eq:lineareq})
constitute a system of infinitely many linear equations with delay, in
the unknowns  $\lattb{{u}}{h}{j}$.

\section{Frequency-dependent dynamical matrix}\label{sec:matrix}

The problem of discussing equations \eqref{eq:lineareq} and solving
them   represents a formidable task. As usual, our study
will be
restricted to the search for \emph{generalized normal modes},
i.e., oscillating modes, possibly including damped and unstable ones. Furthermore, by
factoring the spatial cell dependency in the form of a plane wave, the
translational symmetry of the crystal is exploited to get rid of the
cell index, thus obtaining a parametric dependence on a wavevector
$\mvec{k}$ varying in the Brillouin zone.

So we  substitute  the plane-wave \emph{ansatz}
\begin{equation}
\label{eq:ansatz}
\lattb{u}{h}{j}(t) = \mvec{U}^{\scripts{(j)}} e^{i
  \mvec{k} \cdot \mvec{r}_{\mvec{h}} } e^{- i \omega t}
\end{equation}
into the equations of motion and look at the corresponding equations.
One obviously has 
\begin{align*}
m^{\scripts{(j)}} \lattb{\ddot{u}}{h}{j} &= -\omega^2
m^{\scripts{(j)}}  
\mvec{U}^{\scripts{(j)}} e^{i
  \mvec{k} \cdot \mvec{r}_{\mvec{h}} } e^{- i \omega t} \ ;
\\
 \lattf{h}{j}{coul} &= -  e^{i \mvec{k} \cdot \mvec{r}_{\mvec{h}} }
 e^{- i \omega t} 
 q^{\scripts{(j)}} \sideset{}{'}\sum_{(\mvec{p},s)}
 q^{\scripts{(s)}} \curl \, \curl \, \frac{ \mvec{U}^{\scripts{(j)}} }{
   \abs[\big]{\mvec{x}-\lattb{r}{p}{s}}}
 \Bigg\vert_{\mvec{x}=\lattb{r}{h}{j}} ; \\ 
\lattf{h}{j}{rr} &=
 -i\omega^3 \frac{2}{3} \frac{ \left( q^{\scripts{(j)}} \right)^2 }{ c^3}
  \mvec{U}^{\scripts{(j)}} e^{i  \mvec{k} \cdot \mvec{r}_{\mvec{h}} } e^{- i \omega t} \ .
\end{align*}
and furthermore, as usual in lattice dynamics,
\begin{align*}
\lattf{h}{j}{rep} &= \sum_{s=1}^{n_j} \Bigl( \alpha_{(j)s}
\Pi_{(j)s}^{\parallel} + \beta_{(j)s} \Pi_{(j)s}^{\perp} \Bigr) \cdot 
\bigl(\mvec{U}^{\scripts{\tau_j(s)}} e^{i \mvec{k} \cdot (\mvec{r}_{\mvec{h}}- \mvec{r}_{\mvec{d}_j(s)})} e^{- i \omega t}
 - \mvec{U}^{\scripts{(j)}} e^{i \mvec{k} \cdot \mvec{r}_{\mvec{h}} }
 e^{- i \omega t}\bigl) \\
&= e^{i \mvec{k} \cdot \mvec{r}_{\mvec{h}} } e^{- i \omega t}\sum_{s=1}^{n_j} \Bigl( \alpha_{(j)s}
\Pi_{(j)s}^{\parallel} + \beta_{(j)s} \Pi_{(j)s}^{\perp} \Bigr) \cdot 
\bigg(\mvec{U}^{\scripts{(\tau_j(s))}} e^{- i \mvec{k} \cdot \mvec{r}_{
    \mvec{d}_j(s)} } - \mvec{U}^{\scripts{(j)}} \bigg)  \ .
\end{align*}
For what concerns the dipole terms, instead, one obtains a
qualitatively different contribution, because,  
due to  retardation, there appear terms
 which depends on $\omega$ not simply through the factor $\exp
 (-i\omega t)$.
In fact one has
\begin{align}\label{eq:forzadip}
\lattf{h}{j}{dip} &=
 q^{\scripts{(j)}} \sideset{}{'}\sum_{(\mvec{p},s)} q^{\scripts{(s)}} \curl \,
 \curl \, \frac{ \mvec{U}^{\scripts{(s)}} e^{i
  \mvec{k} \cdot \mvec{r}_{\mvec{p}} } e^{- i \omega t - \frac{\omega}{c}
   \abs{\mvec{x}-\lattb{r}{p}{s}}}}{\abs{\mvec{x}-\lattb{r}{p}{s}}}
 \Bigg\vert_{\mvec{x}=\lattb{r}{h}{j}} \nonumber\\
 &=e^{i \mvec{k} \cdot \mvec{r}_{\mvec{h}} }  e^{- i \omega
   t}q^{\scripts{(j)}} \sideset{}{'}\sum_{(\mvec{p},s)} 
   q^{\scripts{(s)}} 
    \curl \, \curl \, \frac{ \mvec{U}^{\scripts{(s)}} 
     e^{i \mvec{k} \cdot (\mvec{r}_{\mvec{p}}-\mvec{r}_{\mvec{h}}) }e^{- \frac{\omega}{c}
   \abs{\mvec{x}-\lattb{r}{p}{s}}} }
 {\abs{\mvec{x}-\lattb{r}{p}{s}}}
 \Bigg\vert_{\mvec{x}=\lattb{r}{h}{j}} \nonumber\\
 &=e^{i \mvec{k} \cdot \mvec{r}_{\mvec{h}} }  e^{- i \omega t}q^{\scripts{(j)}} 
 \sideset{}{'}\sum_{(\mvec{p},s)} q^{\scripts{(s)}} \curl \, \curl \, 
  \frac{ \mvec{U}^{\scripts{(s)}} e^{- i \mvec{k} \cdot
      \mvec{r}_{\mvec{p}} } e^{- \frac{\omega}{c}\abs{\mvec{x}}}}  {\abs{\mvec{x}}}
 \Bigg\vert_{\mvec{x}=\lattb{r}{p}{j,s}} \ ,
\end{align}
where in the last line we simply replaced the dummy index
$\mvec{h}-\mvec{p}$ by $\mvec{p}$.
So, substituing the above relations into equation (\ref{eq:lineareq}),
after dividing by the common
factor $e^{i \mvec{k} \cdot \mvec{r}_{\mvec{h}} } e^{- i \omega t}$
one gets to the  set of linear equations 
in the unknown $ \mvec{U}^{\scripts{(j)}}$ 
\begin{equation}
\label{eq:normalmodes}
  \begin{split}
    - m^{\scripts{(j)}} \omega^2 \mvec{U}^{\scripts{(j)}} &=
    \sum_{s=1}^{n_j} \Bigl( \alpha_{(j),s} \Pi_{(j),s}^{\parallel} +
    \beta_{(j),s} \Pi_{(j),s}^{\perp} \Bigr) \cdot \bigg(
    \mvec{U}^{\scripts{(\tau_j(s))}} e^{- i \mvec{k} \cdot \mvec{r}_{
        \mvec{d}_j(s)} } - \mvec{U}^{\scripts{(j)}} \bigg) + \\  
    & -q^{\scripts{(j)}} \sideset{}{'}\sum_{(\mvec{p},s)} q^{\scripts{(s)}} \curl\ \curl
    \frac{\mvec{U}^{\scripts{(j)}} }{ \abs{\mvec{x}}}\bigg\vert_{\mvec{x}=\scrlattb{r}{p}{j,s}}  + \\
    & + q^{\scripts{(j)}} \sideset{}{'}\sum_{(\mvec{p},s)} q^{\scripts{(s)}} \curl\ \curl 
    \frac {\mvec{U}^{\scripts{(s)}} e^{- i \mvec{k} \cdot \scrlatt{r}{p}}
      \, e^{ i \frac{\omega}{c} \abs{\mvec{x}}}}  
          {\abs{\mvec{x}}} \bigg\vert_{\mvec{x}=\scrlattb{r}{p}{j,s}} + 
          + i \frac{2}{3} \frac{ \left( q^{\scripts{(j)}} \right)^2 }{ c^3} \omega^3 \mvec{U}^{\scripts{(j)}} .
  \end{split}
\end{equation}
With  $j=1,2,\dots,n$, we have in all $3n$ linear equations, in which
$\mvec{k}$ enters as a parameter and $\omega$ as an unknown. Such
equations  can be
written symbolically in the form
\begin{multline}
- m^{\scripts{(j)}} \omega^2 \mvec{U}^{\scripts{(j)}} = \\
\sum_{s=1}^{n} \bigg[ \hat{\mathcal{P}}_{js}(\mvec{k})  + 
 \hat{\mathcal{C}}_{js}  +
 \hat{\mathcal{D}}_{js}(\mvec{k},\omega) \bigg] \cdot \mvec{U}^{\scripts{(s)}}  +
i \frac{2}{3} \frac{ \left( q^{\scripts{(j)}} \right)^2 }{ c^3} \omega^3 \mvec{U}^{\scripts{(j)}} \; ,
\end{multline}
having denoted by $\mathcal{P}= \{ \hat {\mathcal{P}}_{js}\}$, 
$\mathcal{C}=\{ \hat{\mathcal{C}}_{js}\}$,
$\mathcal{D}=\{ \hat{\mathcal{D}}_{js}\}$ 
respectively 
the matrix of the  short-range repulsive forces, that of the Coulomb
forces, and that of the 
dipole ones. In short, we can also write the equations for the normal 
modes in the form
\begin{equation}
\label{eq:symbolicalseceq}
- m^{\scripts{(j)}} \omega^2 \mvec{U}^{\scripts{(j)}} =
\sum_{s=1}^{n} \hat{\mathcal{A}}_{js}(\mvec{k},\omega) \cdot
\mvec{U}^{\scripts{(s)}} 
\end{equation}
which involves  a \emph{dynamical matrix} $\mathcal{A}$ in a way
apparently similar to that occurring
in  the case of the ``instantaneous'' Coulomb model.
A sharp difference is however that
the dynamical matrix $\mathcal{A}$ now
depends on the \emph{unknown} $\omega$, besides on the wavevector parameter
$\mvec{k}$. Such a dependency is obviously contained in the dipole
part.  \emph{This is a peculiar and remarkable consequence of
retardation}, which entails that equation \eqref{eq:symbolicalseceq} does \emph{not} give
rise to a standard secular equation. Denoting by 
$\mathcal{M}=\mathrm{diag}(m_1,\dots,m_n)$ the
mass matrix, the   analogue of the secular equation presently takes the form
\begin{equation}   
\label{eq:secular}
\det \Big( \omega^2 \mathcal{M} + \mathcal{A}(\mvec{k},\omega) \Big)
= 0 .
\end{equation}
This makes the dispersion relations more difficult to compute. 
Due to  the $\omega$-dependency of the
dynamical matrix, the
solutions cannot be worked out through standard diagonalization 
methods of  linear algebra, and a more general numerical algorithm is required.

\begin{figure*}[ht]
  \centering 
  \includegraphics[scale=0.725]{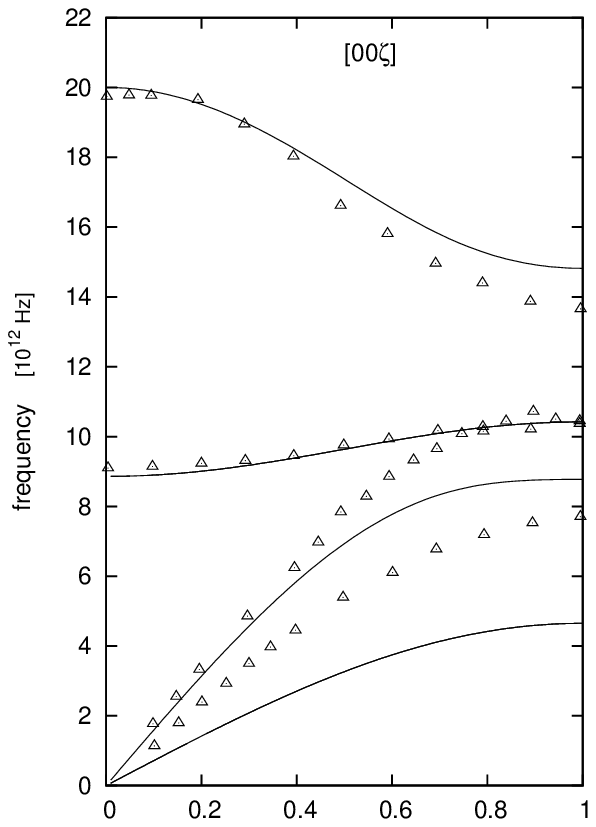}
  \includegraphics[scale=0.725]{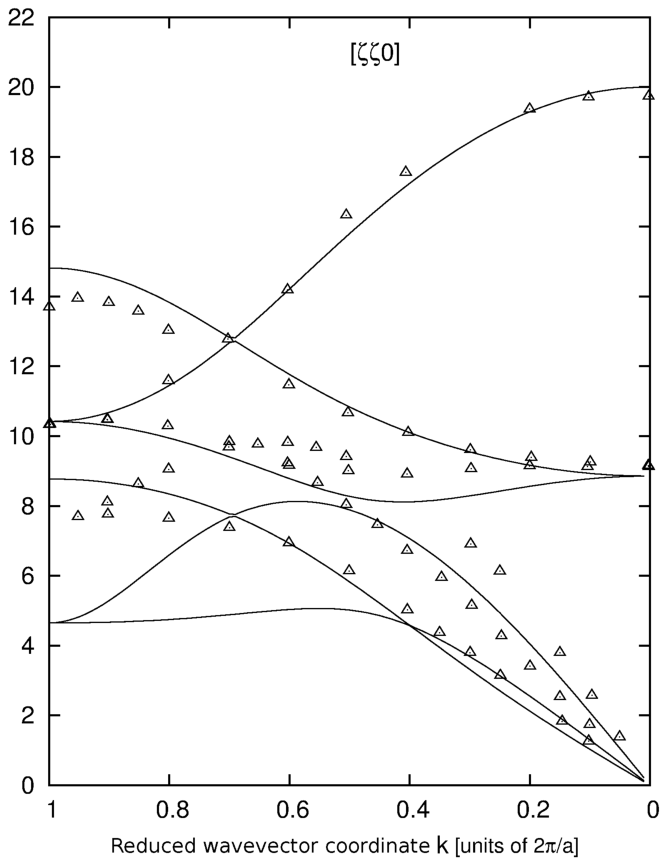}
  \includegraphics[scale=0.725]{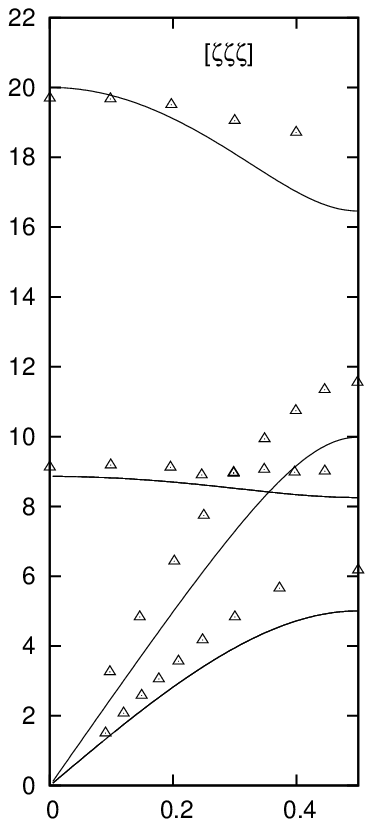}
  \caption{Dispersion relations   along three directions of high symmetry
    in LiF. Solid  curves are the theoretical predictions,  and
    triangles   the experimental data taken from 
\cite{exper} }\label{fig:1} 
\end{figure*}
Moreover, we see that the $\omega$-dependency of the entries of
$\mathcal{A}$ is by no means simple; actually it is not algebraic, neither 
can it be expressed in terms of elementary functions. Thus, it is
not even possible to foresee how many solutions,  for each value of
$\mvec{k}$ do exist. The number of branches and their topology may vary in a
substantial way  with respect to the ``mechanical or instantaneous'' 
case, which displays
$3n-3$ optical branches and $3$ acoustic ones. This is
indeed what makes the existence of polaritonic branches possible.

One sees  that the instantaneous Coulomb limit case is formally obtained
by taking the limit $\omega \to 0$  (or equivalently $c \to \infty$) 
in the dipole field in \eqref{eq:normalmodes}. In such a limit, the sum
of the Coulomb term and of the
dipole one gives simply an additional ``mechanical'' term, completely
analogous to the molecular repulsive one,  the only difference being
that the long range
of the Coulomb forces makes a resummation procedure
necessary. However, it turns out that
 this limit misses most of the physics
involved in the region about the e.m. dispersion curve $\omega = c k$, 
which is the primary goal of the present discussion.

The sums over the lattice, which define the electric fields in
\eqref{eq:normalmodes}, are obviously ill-defined, due to the long
range of the e.m. interactions. A precise meaning is assigned to them
through
the well-known \emph{Ewald's summation} procedure\cite{ewald}. 
This  transforms each
conditionally convergent series into the sum of two rapidly convergent ones,
which account for the ``short--distance'' part
 and the ``long--\-distance'' part of the
interactions respectively. The use of such a method is crucial, for it provides a
deep insight into the physical aspect of the problem. It 
turns out that only the term
describing long-distance interactions is substantially modified by the
$\omega$-dependency (i.e., by retardation), and that such dependency
is important only in the region about the e.m. dispersion curve $\omega = ck$.

The details of the Ewald's summation method can be found either in his
original paper \cite{ewald} or in more recent works
(e.g., \cite{newewald}). A  compact exposition can also be found in
the appendix of \cite{chaos}.  A slight generalization of the method 
had to be devised in
order to deal with the dipole fields. This is discussed in Appendix
\ref{app:ewald}. 
 It turns
out that, once the  Ewald resummation has been performed, 
the dipole part of the 
dynamical matrix  takes, for  $j\neq l$, the   structure 
\begin{multline}
\label{eq:ewalddipolefar}
\hat{\mathcal{D}}_{jl} = q^{\scripts{(j)}} q^{\scripts{(l)}} 
\Bigg\{ \frac{4\pi}{\abs{V_c}}  \sum_{\mvec{m} \in \mathbb{Z}^3}
\frac{e^{ - \frac{ 1}{4\delta'^2} \big( \abs{\latt{q}{m}-\mvec{k}}^2 -
    \frac{\omega^2}{c^2} \big) }} { \abs{\latt{q}{m}-\mvec{k}}^2 -
  \frac{\omega^2}{c^2} } \hat{\mathcal{R}}\big[e^{+ i (\latt{q}{m}-\mvec{k})
    \cdot
    \mvec{x}}\big]\bigg\vert_{\mvec{x}=\lattb{b}{}{j,l}} \\
 + \sum_{\mvec{h} \in \mathbb{Z}^3} e^{ - i \mvec{k} \cdot
  \latt{r}{h}} \frac{2}{\sqrt{\pi}} \int_{\delta'}^{+\infty} d\eta \,
e^{\frac{\omega^2}{4c^2}\frac{1}{\eta^2}} \hat{\mathcal{R}} \Big[ e^{-
    \abs{\mvec{x}}^2 \eta^2 }
  \Big]\bigg\vert_{\mvec{x}=\lattb{r}{h}{j,l}} \Bigg\} \ ,
\end{multline} 
where the first term describes
the field associated to the far charges, while the second one accounts for
the short-range part of the interactions. The notations
are as follows:
$\latt{r}{h}$ runs over the direct lattice, $\latt{q}{m}$ runs over
the reciprocal lattice; $\lattb{b}{}{j}$  for $j=1,\dots,n$ are the
positions of the ions within a unit cell
($\lattb{r}{h}{j}=\latt{r}{h}+\lattb{b}{}{j}$); ${V_c}$ is the
region of a cell and $\abs{V_c}$ its volume. Furthermore,
$\hat{\mathcal{R}}$ is the 
matrix form of the differential operator $\curl \ \curl$: for a
scalar field $f$ one has 
\[ 
\hat{\mathcal{R}}[f(\mvec{x})] = \hat{H} [f(\mvec{x})] -
\Delta[f(\mvec{x})] \hat{\mathbb{I}}_3 
\] 
where $\hat{H}[f]$ is the Hessian of $f$ (the matrix of its second 
derivatives), $\Delta$ the Laplacian and
$\hat{\mathbb{I}}_3$ the identity matrix.  

For $j=l$ the matrix element
has  just the same form, apart from the fact that the 
unphysical self-interaction term must be subtracted. 
Here a very interesting fact occurs: such a subtracted term gives rise
to a non-hermitian part for the dipole term matrix, that
is \emph{exactly canceled}  by the radiation reaction term (the last one
in \eqref{eq:normalmodes}, which is clearly non-hermitian, too). 
Indeed, as shown in Appendix~\ref{app:WF}, for all $j,s=1,\dots,n$ one has
\[
 \frac{1}{2} \bigg( \hat{\mathcal{D}}_{js}(\mvec{k},\omega) - \hat{\mathcal{D}}_{sj}^{\dagger}(\mvec{k},\omega) \bigg) + i \frac{2}{3} \frac{q^{\scripts{(j)}}}{c^3} \omega^3 \delta_{js} \hat{\mathbb{I}}_3 = 0 .
\]
Hence, the complete dynamical matrix $\mathcal{A}$ is hermitian, 
so that only real
frequencies are allowed. We thus see that the radiation reaction force
is vital to the stability of the lattice: a nonvanishing imaginary part in
the frequency would correspond to damped or unstable
oscillations. This a priori unexpected cancellation, first realized by 
Oseen\cite{oseen}, and then rediscovered in \cite{cg1,marino1}, 
might seem ``accidental'' at first sight. The proof of such a
cancellation, given in
appendix~\ref{app:WF}, shows its  deep meaning,
first pointed out by  Wheeler and Feynman\cite{wf}.

The \emph{splitting parameter} $\delta'$ in~\eqref{eq:ewalddipolefar}, 
involved in Ewald's method, 
is arbitrary, to be chosen so that 
both series converge rapidly. If it is taken of the order of the
inverse first-neighbor interatomic distance, one sees that
\[  
\frac{\omega}{2 c \eta} \lesssim 10^{-5} \qquad \text{for all} \;
\eta > \delta' \ ,
\] 
if one takes $\omega$ in the range of the typical frequencies of
crystal dynamics, i.e., $\omega\simeq 10^{13}$ Hz.
Hence we may safely take $\omega=0$ in the corresponding
exponential in the second term: the $\omega$--dependency of the
short--distance part of the interactions is negligible, so that they can be considered
as instantaneous. Instead, the first term, which describes the
long--distance part of the interactions, is strongly frequency-dependent:
the term $\mvec{m}=\mvec{0}$ has in fact a pole at $\omega=ck$,
i.e., along the e.m. dispersion line. 

This is the main mathematical effect
of retardation: the part of the dynamical matrix which describes
the interaction with the far charges strongly depends on the frequency
$\omega$ near the e.m. dispersion curve $\omega=ck$. Hence, very
different predictions are expected with respect to the instantaneous 
Coulomb model. In the following, we show that such a feature leads indeed
to retrieve the polariton curves.

\section{Polaritons  in a rock salt lattice}

At this point  we are no longer able to proceed any further on a general
discussion. In order to check  concretely the predictions of the model,
a numerical study is needed, and so we concentrate on a specific
crystal structure. In view of a  comparison with experimental data, we choose
the \emph{rock salt} structure, which is in fact very simple and is
shared by the most common alkali halides. Concretely, the comparison
with the experimental data was  performed for the case of Lithium
Floride, LiF. 
So, we first have to explicitly write down
the dynamical matrix $\mathcal{A}$ in general, which   
is done in appendix \ref{app:dynmatrix}. Then, the
expressions thus found have to be specialized to the case 
of the rock salt lattice.

In order to solve numerically the generalized secular equation
\eqref{eq:secular}, with the dynamical matrix $\mathcal{A}$  now
computed for a generic rock salt lattice, we devised a very general
and straightforward algorithm. The overall idea is the following. The
wavevector 
$\mvec{k}$ is let run along certain directions from the zone center
to the zone boundary, and the frequency $\omega$ is let vary in a
suitable range. Then, considering both $\mvec{k}$ and $\omega$ as
parameters,  the
dynamical matrix is numerically evaluated, and a generalized
diagonalization is performed, by determining the six values 
$\lambda_i(\mvec{k},\omega)$ such
that $\det \Big( \lambda_i \mathcal{M} + \mathcal{A}(\mvec{k},\omega) \Big)
= 0$. Eventually, leaving $\mvec{k}$ only as a parameter,
the roots of the equations
\begin{equation} 
\label{eq:numerical}
\lambda_i(\mvec{k},\omega) = -\,  \omega^2
\end{equation}
are numerically determined for each $i=1,\ldots,6$. This yields the admissible
excitation frequencies of the lattice for the given wavevector
$\mvec{k}$. In particular, as previously mentioned,   
the pole in the matrix elements along the
line $\omega=ck$ gives rise to a corresponding pole
for certain eigenvalues (as functions of $\omega$), and hence the
number of solutions of \eqref{eq:numerical} doubles: at variance with
the instantaneous Coulomb model, we have more than $3n$
branches.
For values of $\mvec{k}$ and $\omega$ far from the line $\omega=ck$ 
the whole procedure is actually redundant, because
the instantaneous approximation is perfectly suitable. 

Now, the dispersion curves thus found still depend on the three free
parameters entering the model, which are determined by a best fit
with the experimental phonon  curves. 
The best fit thus
obtained is exhibited in   figure \ref{fig:1}. 
In table \ref{tab:valuesLiF} the numerical values employed for 
the parameters entering the model are collected.

The three panels give $\omega$ versus $k$ for the three high--symmetry
directions of $\vect k$ $(0,0,1)$ (left), $(1,1,0)$ (center) and $(1,1,1)$
(right) for the whole Brillouin zone. The triangles are the
experimental values taken from \cite{exper}, while the continuous
lines give the theoretical curves. The global  agreement over the
whole Brillouin zone seems to us to be  fairly good, in consideration of
the simplicity of the model, and of the small number (three) of free 
parameters. In fact, one can notice that the fit is not so good for the
acoustic branches, especially for high $k$. A better fit over the
whole Brillouin zone was actually  obtained in the paper \cite{exper}, using
  second--nearest--neighbor short range forces, in a model
involving seven free parameters. So, presumably, an analogous better
fit could have been obtained by us too by making recourse to a
more refined model for the short range forces.  
However, we did not care for this, because our main
goal is to exhibit the occurrence of
polaritons in the   simplest possible way.

\begin{table}
\centering
\label{tab:valuesLiF}
\begin{tabular}{cc}
\toprule
Parameter & Value \\
\midrule
$\alpha$  & $-0.19$  \\
 $\beta$   &  $ + 0.028$        \\
$q$                &    $0.6$             \\
  $a$                   &     $ 4.02 $     \\
   $m^+$           &     $0.005$     \\
 $m^-$           &     $0.0136$     \\
\bottomrule
\end{tabular}
\caption{Parameters for LiF (units: \AA, $10^{-13} s$, $e$). The first 
three ones (the force constants $\alpha,\beta$ and the effective
charge of both ions $q$) have been obtained by fitting the
experimental phonon dispersion curves taken from \cite{exper}. For the
lattice parameter $a$ and the masses $m^+,m^-$ the commonly accepted
values have been employed.}
\end{table}

\begin{figure}[ht]
 \centering
  \includegraphics[scale=0.9]{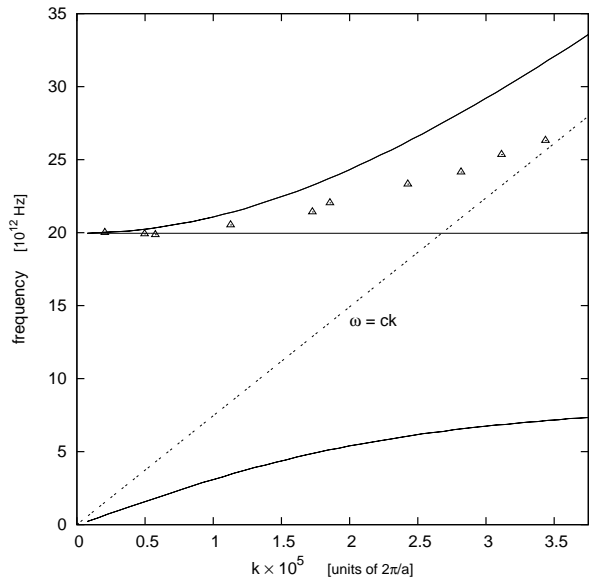}
  \caption{Zoom of the central panel of  figure \ref{fig:1} 
for small $k$, exhibiting 
    polaritonic branches. Solid curves are the theoretical
    predictions, dashed line is the curve $\omega =ck$, while
    triangles  are the experimental data at $420 \, K$, taken 
    from  \cite{osanohikamoto}. Acoustic branches superimpose
     to the axis $\omega=0$}\label{fig:2}
\end{figure}

Now,  the polaritonic branches cannot be seen in
figure~\ref{fig:1}
because they are  squeezed along  the ordinate axis. They are
exhibited in figure~\ref{fig:2}, in which 
a zoom of the left panel of figure~\ref{fig:1} is
performed, by enlarging by a factor $10^5$ the axis
of the absciss\ae. 

As expected, near the line
 $\omega=ck$,  the  number and 
 the topology of the branches are
drastically different with respect to those  of the instantaneous 
approximation, and presents   a  pattern displaying  polaritonic 
curves.
As  the acoustic branches are now squeezed on the absciss\ae\
axis, one should look at the optical branches, which in figure
\ref{fig:1} intersect the axis of the ordinates at  $\nu \approx 9\ 10^{12}$ Hz
(the two transverse ones) and at $\nu\approx 20\ 10^{12}$ Hz (the
longitudinal one). One sees that the two transverse branches  actually go
to zero, with a slope around $c/2.27$, 
whereas there appear two new branches,
actually degenerate, which start
from the fundamental frequency of the longitudinal--modes
branch at $k=0$
and are asymptotic to the e.m. dispersion line $\omega = c
k$.
Namely,  the transverse optical branch (which actually
represents the two transverse--mode curves) \emph{splits} into a
lower branch, which approaches zero, and an upper one.  
Such a phenomenon is a peculiar
effect of retardation, which strongly couples the radiation field
of far charges to the vibrational modes. 

The
longitudinal--modes branch, instead, is untouched by  retardation.
This is due to the fact that the electromagnetic waves, being 
 transverse, do not couple with longitudinal
modes. Indeed, the only terms in~\eqref{eq:ewalddipolefar}
which are affected by  retardation, i.e., the terms $\mvec{m}=\mvec{0}$
of the sums over the reciprocal lattice, contain
a projection onto the  plane  orthogonal to the wavevector $\mvec{k}$,
which yields zero when applied to longitudinal displacements 
$\mvec{U}^{\scripts{(l)}}$.

In figure \ref{fig:2} are reported also (triangles) the experimental
values taken from the work \cite{osanohikamoto},
and  one sees that
the agreement   is  pretty good for small $k$, while it is not so
good for the asymptotic behavior
at  larger $k$. Actually we found that the agreement in  the small $k$
region  depends very 
critically on the value of  the effective charge, which  had to be
 carefully chosen.
Instead, the partial discrepancy
in the high $k$ region appears to be rather
due   to a deficiency 
of the model itself, inasmuch as 
the ions are dealt with as point charges without any internal structure.
In order to appreciate this fact one should notice that,
 as better discussed in the following section, the dispersion
curves for the lattice vibrations coincide with the
dispersion curves of the e.m. field  propagating inside  the crystal.
In other terms,  the slope of the lower branch in the
low--frequency region of figure \ref{fig:2}   should coincide with
the speed of e.m. field propagation inside the crystal, i.e., with
\[
c/\sqrt{\varepsilon} \, , 
\] 
where $\varepsilon$ is the
\emph{static dielectric constant} of the medium.
In the usual macroscopic treatments of
electromagnetism and optics, such a constant  is a
phenomenological parameter related to the polarizability of the system. 
Here, instead, it arises in a natural way
as a consequence of the microscopic dynamics, and its value is deduced
from the geometrical structure of the lattice and from the physical
parameters which characterize the ionic crystal under consideration
(masses, charges and the repulsive force constants). In our case  we
found $\varepsilon \approx 5.3$, while experimentally one finds
$\varepsilon\approx 9$. We  expect that 
this discrepancy should be 
attributed to the approximation of considering   the ions as point
particles, thus neglecting   polarizability. 
Now, we expect  that this approximation should  be responsible also
for the discrepancy concerning the  upper transverse polariton high-frequency
slope, which here is $c$, while being   experimentally  smaller.
Indeed the modification of the speed of the e.m. field at the optical 
frequencies  should be due to the interactions with the electrons, which
is neglected in the present model.

We also point out that  between the upper and the lower
transverse branches there occurs  a frequency gap, which should 
correspond to the
frequency of the infrared residual rays.

\section{Deduction of macroscopic optics}\label{sec:optics}

So far we have deduced, calculated and discussed the mere
\emph{vibrational} properties of ionic lattices, by determining their
normal modes  when the electromagnetic forces are taken into account. 
In this section we
show how such microscopic lattice dynamics gives rise to a macroscopic
propagation of electromagnetic fields across the lattice, in agreement with the
laws of macroscopic optics.

The field $\mvec{E}(\mvec{x},t)$ propagating across the lattice is obtained in the linear
approximation by evaluating the superposition of the variable electric
fields generated by the oscillations of the dipoles at each lattice
point, which gives
\begin{equation*}
\mvec{E}(\mvec{x},t) = \lattser{h}{j} \curl \  \curl\ \bigg(
q^{\scripts{(j)}} \frac{\lattb{u}{h}{j}\big( t - \frac{1}{c}
  \abs{\mvec{x}-\lattb{r}{h}{j}}
  \big)}{\abs{\mvec{x}-\lattb{r}{h}{j}}} \bigg) .
\end{equation*}

Let us now suppose that the ions are oscillating according to a normal mode,
identified by $\omega$ and $\mvec{k}$: the law of motion of each dipole is thus
\[
\lattb{u}{h}{j}(t) = \lattb{U}{}{j} e^{i \mvec{k} \cdot \latt{r}{h}}
e^{-i\omega t},
\]
for some definite amplitude vectors $\{
\lattb{U}{}{j}(\mvec{k},\omega) \}_{j=1,\dots,n}$. The field is then
\begin{multline*}
\mvec{E}(\mvec{x},t) = \\ e^{-i\omega t} \curl \ \curl\
\Bigg[\sum_{j=1}^{n} q^{\scripts{(j)}} \Bigg( \sum_{\mvec{h} \in
    \mathbb{Z}^3} e^{i \mvec{k} \cdot \latt{r}{h}} \frac{e^{i
      \frac{\omega}{c}\abs{\mvec{x}-\lattb{r}{h}{j}}}}{\abs{\mvec{x}-\lattb{r}{h}{j}}}
  \Bigg) \lattb{U}{}{j} \Bigg],
\end{multline*}
or equivalently
\begin{equation}
\label{eq:fieldpropagation}
\mvec{E}(\mvec{x},t) = e^{-i\omega t} \curl \ \curl\
\Bigg[\sum_{j=1}^{n} q^{\scripts{(j)}} \,
  \Psi(\mvec{x}-\lattb{b}{}{j}) \, \lattb{U}{}{j} \Bigg] \ ,
\end{equation}
where we have introduced
\[
\Psi(\mvec{x}) \definedas \sum_{\mvec{p}\in \mathbb{Z}^3} e^{ + i
  \mvec{k} \cdot \scrlatt{r}{p} } \frac{e^{i \frac{\omega}{c}
    \abs{\mvec{x}-\latt{r}{p}}}}{\abs[\big]{\mvec{x}-\latt{r}{p}}}.
\]
As shown in Appendix~\ref{app:ewald}, we can now carry on the
manipulation on $\Psi(\mvec{x})$, 
leading to a sum of rapidly convergent series,
namely
\begin{equation}\label{eq:splitphi}
\begin {split}
\Psi(\mvec{x}) & = \Psi_1(\mvec{x}) + \Psi_2(\mvec{x}) \definedas 
\frac{4 \pi}{\abs[\big]{V_c}} \sum_{\mvec{m} \in \mathbb{Z}^3} \frac{e^{-
    \frac{1}{4\delta'^2} \big( \abs[\big]{\latt{q}{m} - \mvec{k}}^2 -
    \frac{\omega^2}{c^2} \big) }}{\abs[\big]{\latt{q}{m} - \mvec{k}}^2
  - \frac{\omega^2}{c^2} } e^{i (\mvec{k} - \latt{q}{m} ) \cdot
  \mvec{x} } + \\ 
& + 
\sum_{\mvec{h} \in \mathbb{Z}^3} e^{ i \mvec{k} \cdot \latt{r}{h}}
\frac{2}{\sqrt{\pi}} \int_{\delta'}^{+\infty}
e^{\frac{\omega^2}{4c^2}\frac{1}{\eta^2}} \, e^{-
  \abs{\mvec{x}-\latt{r}{h}}^2 \eta^2 } d\eta \ .  
\end{split}  
\end{equation}
Exploiting the algebraic identity
\[
a^2 - b^2 = (a \pm i b)^2 \mp 2 i a b
\]
in the exponent of the integrand, it is convenient to rewrite the
second part $\Psi_2(\mvec{x})$ of $\Psi$ in the  remarkable form
\begin{equation}
\label{eq:microscopicfield}
\Psi_2(\mvec{x}) e^{-i\omega t} =  \sum_{\mvec{h} \in \mathbb{Z}^3}
\bigg( e^{ i \mvec{k} \cdot \latt{r}{h}} \frac{2}{\sqrt{\pi}}
\int_{\delta'}^{+\infty} e^{\big( \frac{\omega}{2c}\frac{1}{\eta} - i
  \abs{\mvec{x}-\latt{r}{h}} \eta \big)^2} d\eta \bigg)  e^{i \big(
  \frac{\omega}{c}\abs{\mvec{x}-\latt{r}{h}} - \omega t \big)}\ .
\end{equation}
This expression highlights that, for a  not
too small splitting parameter $\delta'$ (say, comparable to the inverse interatomic first-neighbor
distance), $\Psi_2$ is roughly the superposition of spherical waves
coming from the neighboring sites and propagating at speed
$c$. \\ The function $\Psi_1(\mvec{x})$ can instead be written as
\begin{equation}
\label{eq:macroscopicfield}
\Psi_1(\mvec{x}) e^{-i\omega t} =  \Bigg(\frac {4 \pi} {\abs[\big]{V_c}}
\sum_{\mvec{m} \in \mathbb{Z}^3} \frac{e^{- \frac{1}{4\delta'^2} \big(
    \abs[]{\latt{q}{m} - \mvec{k}}^2 - \frac{\omega^2}{c^2} \big)
}}{\abs[\big]{\latt{q}{m} - \mvec{k}}^2 - \frac{\omega^2}{c^2} } e^{-
  i \latt{q}{m} \cdot \mvec{x} } \Bigg) e^{i (\mvec{k} \cdot \mvec{x}
  - \omega t)}\ .
\end{equation}
The term in brackets is a smooth quasiperiodic function over the direct
lattice. Thus  $\Psi_1$ looks like a \emph{plane
  wave} characterized by the phonon wavevector $\mvec{k}$ and
the frequency $\omega$, hence propagating at the speed
\begin{equation}
\label{eq:refractionindex}
v = \frac{\omega}{\abs{\mvec{k}}} \definedasright \frac{c}{n_b
  (\mvec{k})}\ .
\end{equation}
The refraction index $n$ depends on the considered normal mode,
identified by a wavevector $\mvec{k}$ plus a branch label
$b=1,2,\dots$,  corresponding to the specific branch considered.

Substituting such expressions of $\Psi_1$ and $\Psi_2$ into
\eqref{eq:fieldpropagation} and computing the effect of $\curl \
\curl$ on them, we find the explicit form of the propagating field. We
see that the term originating from \eqref{eq:microscopicfield} varies
over a microscopic scale (i.e., one comparable to the linear dimensions of
a primitive cell), whereas the term resulting from
\eqref{eq:macroscopicfield} is much ``smoother''. If we consider  that
a reasonable measuring instrument should be necessarily
\emph{macroscopic}, hence much larger than the atomic scale, then only
the latter field should be observable. Therefore, we name the former
\emph{microscopic field}, and the latter \emph{macroscopic
  field}. Such a distinction is significant at macroscopic scales,
i.e., in the usual frame of  the elementary (phenomenological) 
treatments of optics.

This result means that at least a relevant part of the infrared dispersion
phenomenology in ionic crystals can be deduced from the \emph{vibrational}
dispersion relations: indeed through equation \eqref{eq:refractionindex} the
refraction index can be expressed as a function of frequency. The
interaction between radiation and matter (or \emph{phonon-photon
  coupling}, in the quantum picture) can be interpreted in terms of
such relations.

\section{Final considerations}

So,  by studying  a microscopic classical model of 
an ionic crystal with the ions dealt with as point charges, 
we have shown that the retarded action of the \emph{far} ions is 
responsible for the
splitting of the dispersion curves about the e.m. dispersion line
$\omega=ck$, i.e., for the existence of polaritons in ionic crystals.

A fairly good quantitative agreement between the theoretical
polaritonic curves and the experimental data for  Lithium Fluoride
 is obtained directly from  microscopic dynamics,
without further \textit{ad hoc}
assumptions or the use of any fitting parameters, apart from the three
parameters (effective charge of the ions and two constants entering
the repulsive forces) related to the non--retarded part
of the problem. In any case, polaritons come out automatically in virtue of
retardation, without the need of introducing any new parameter, 
once  the instantaneous part of the problem has been
settled. For what concerns the choice of the free parameters, one may notice  
that they could  also be determined from experimental data not related
to the dispersion curves, as for example  thermodynamic quantities such as
the internal energy or  data on the infrared absorption (see \cite{born}).

One may now ask  whether an  explanation of dispersion can in
some analogous way be given also for other types of   crystal
insulators. The simplest model is obtained by considering a 
 lattice of pure dipoles. Such a model was already studied
in the work \cite{marino1}, and a  phenomenon analogous to that of  existence
of polaritonic curves was observed.

We finally add now a comment concerning the treatment of the problem
given in the book of Born and Huang\cite{{bornhuang}}.  We  already
mentioned that such authors predicted the existence of polaritons
in the first part of the book,  where the problem  is discussed
in terms of macroscopic  polarization fields. It seems however
 that a  proof  
is lacking in the second part of the book,
which is devoted to a microscopic discussion of the problem. 
Apparently this is due to the fact that, in discussing  the
secular equation, the authors  do not introduce explicitly $\omega$ as an
unknown of the problem, limiting themselves to introduce the \emph{ansatz}
$\omega=(c/n) k$. This entails  that  the upper polaritonic
branch cannot be detected. Moreover, at page 334, they explicitly 
say that \emph{``The last term}
(i.e., the retarded one) \emph{can be ignored''}.

\vskip 1.em
\noindent
\textbf{Acknowledgement}:
We thank Giuseppe Pastori Parravicini. Having read the papers  \cite{cg1}
and \cite{marino1}, where retardation was taken into account 
in microscopic models involving   
internal dipoles only, he  suggested  that polaritons may be proven 
to exist by analogous methods,
 if one considers a model involving the displacements of the ions.

\appendix

\section{The Wheeler--Feynman identity and the stability of the lattice} 

\label{app:WF}

In discussing the secular equation,
it was already pointed out that the  term \eqref{eq:rr}, due
to the radiation reaction force entering the Newton equation for each ion,
exactly cancels the non-hermitian part of the dynamical matrix, so
that one is left with a hermitian dynamical matrix. This has the
consequence that
only real frequencies $\omega$ (i.e. stable non-damped oscillations) are
allowed, so that the lattice can \emph{exist}. 
As mentioned in \cite{cg1}, such a cancellation was first
pointed out by Oseen\cite{oseen} in the year 1916.
 Note that a priori such a cancellation is by no means evident nor obvious:
when radiation effects are taken into account, one might expect that
non-trivial energy exchanges occur, and damped oscillations may arise
due to uncompensated energy losses. It should be noted that, if
$\omega$ has a positive imaginary part, the dipole term series
describing the retarded interaction diverges.  The actual role of the
radiation reaction force deserves thus to be pointed out: its inclusion
in the equations of motion is vital in order to make  the lattice stable,
and in allowing for the existence itself of a dispersion relation; omitting this
term would lead to a substantial inconsistency of the model.

It might seem that the Oseen cancellation occurring for the non-hermitian
part of the dynamical matrix, arises ``accidentally''. Here we provide
a proof,  more 
significant than the straightforward computation implicitly
carried out in the text. The present proof
shows that the origin of the cancellation is actually deeper,
and can ultimately be ascribed to the symmetry of electrodynamics with
respect to time inversion.

In fact, we prove here that the  present model of ionic crystal   satisfies the
following identity, first proposed by Wheeler and Feynman\cite{wf}: 
\begin{equation}
\label{eq:wheelerfeynman}
\sum_k \Big( F_{\text{ret}}^k(\mvec{x},t) -
F_{\text{adv}}^k(\mvec{x},t) \Big) = 0 ,
\end{equation}
where the summation index $k$ runs over all the charged particles of
the system and $F$ is the e.m. field tensor.\footnote{Alternatively,
  we might say that the crystal has the property of being
 a Wheeler-Feynman \emph{complete    absorber}. }
This evidently points out the symmetrical role played by the retarded and
the advanced solutions of the Maxwell equations.
 We will then show
that this identity actually  implies the Oseen cancellation.

First of all, it is clear that the zeroth-order Coulomb fields
trivially satisfy the above identity, because they are independent of
time. In addition, recall that we are neglecting magnetic fields, which give
second-order effects. Thus, verifying \eqref{eq:wheelerfeynman}
amounts to showing that
\begin{equation}
\label{eq:wf2}
 \lattser{p}{s} \Big(
 \mvec{E}_{\mvec{p},s}^{{\ord{1}},\text{ret}}(\mvec{x},t) -
 \mvec{E}_{\mvec{p},s}^{{\ord{1}},\text{adv}}(\mvec{x},t) \Big) = 0
\end{equation}
or, substituting the normal mode \emph{ansatz} \eqref{eq:ansatz},
\begin{equation}
\label{eq:wfidnm}
0 = e^{- i \omega t} \sum_{s=1}^{n} q^{\scripts{(s)}} \curl \ \curl 
\Bigg[
  \sum_{\mvec{p} \in \mathbb{Z}^3} e^{- i \mvec{k} \cdot
    \mvec{r}_{\mvec{p}} } \Bigg( \frac{ e^{i \frac{\omega}{c}
      \abs[]{\mvec{x}-\scrlattb{r}{p}{s}}}}{\abs[\big]{\mvec{x}-\lattb{r}{p}{s}}}
  - \frac{ e^{- i \frac{\omega}{c}
      \abs[]{\mvec{x}-\scrlattb{r}{p}{s}}}}{\abs[\big]{\mvec{x}-\lattb{r}{p}{s}}}
  \Bigg) \lattb{U}{}{s} \Bigg]
\end{equation}
This is immediately seen if one remarks  the   the only difference that
shows up when considering the advanced fields occurs in the spherical-wave
term, whose direction is inward rather  than outward.  

In order to
prove the Wheeler-Feynman identity in the form~\eqref{eq:wfidnm}, we
rewrite the term in square brackets as a sum
over the reciprocal lattice. Using the distributional
identities (see e.g. \cite{tikhonov})
\begin{gather*}\label{eq:sphericalwave}
\frac{e^{\pm i \alpha x}}{x} = 4 \pi \lim_{\epsilon \to 0^+}
\int_{\mathbb{R}^3} dm(\mvec{k}) \frac{e^{i \mvec{k} \cdot
    \mvec{x}}}{\abs{\mvec{k}}^2 - (\alpha^2 \pm i \epsilon)} \, , \\
\label{eq:periodicdelta}
\sum_{\mvec{h} \in \mathbb{Z}^3}\delta(\mvec{x} - \latt{r}{h}) =
\frac{1}{\abs{V_c}} \sum_{\mvec{m} \in \mathbb{Z}^3} e^{i
  \scrlatt{q}{m} \cdot \mvec{x}} \, ,
\end{gather*}
 one gets
\begin{equation*}
  \begin{split}
    \sum_{\mvec{p} \in \mathbb{Z}^3} & e^{- i \mvec{k} \cdot
      \mvec{r}_{\mvec{p}} } \Bigg( \frac{ e^{i \frac{\omega}{c}
        \abs{\mvec{x}-\scrlattb{r}{p}{s}}}}{\abs[\big]{\mvec{x}-\lattb{r}{p}{s}}}
    - \frac{ e^{- i \frac{\omega}{c}
        \abs{\mvec{x}-\scrlattb{r}{p}{s}}}}{\abs[\big]{\mvec{x}-\lattb{r}{p}{s}}}
    \Bigg) = \\ 
    &\frac{4 \pi}{\abs[\big]{V_c}} \sum_{\mvec{m} \in \mathbb{Z}^3}
    \lim_{\epsilon \to 0^+}   \Bigg( \frac{e^{i ( \latt{q}{m} - \mvec{k})
        \cdot (\mvec{x} - \scrlattb{b}{}{s}) }}{\abs[\big]{\mvec{k} -
        \latt{q}{m}}^2 - \big(\frac{\omega^2}{c^2} + i \epsilon\big)} -
    \frac{e^{i (\latt{q}{m} - \mvec{k}) \cdot (\mvec{x} -
        \scrlattb{b}{}{s}) }}{\abs[\big]{\mvec{k} - \latt{q}{m}}^2 -
      \big(\frac{\omega^2}{c^2} - i \epsilon\big)}\Bigg) = 0.
  \end{split}
\end{equation*}
So \eqref{eq:wfidnm} is proven.

We finally show that the Wheeler-Feynman identity leads somewhat
directly to \emph{Oseen identity}, i.e., to the cancellation  between the non-hermitian part
of the dipole term matrix and the radiation reaction term in the
linear equations of motion.

To this end we let $\mvec{x}$ approach a lattice site $\lattb{r}{h}{j}$,
and rewrite \eqref{eq:wfidnm} by separating the contribution of the
$(\mvec{h},j)$ ion from that of all the other ones, thus getting
\begin{multline*}
0 = 
 \frac {e^{- i \omega t}}2 \ \curl \ \curl
\Bigg\{ \ \,  \sideset{}{'}\sum_{(\mvec{p},s)} 
\bigg[ q^{\scripts{(s)}} e^{i \mvec{k} \cdot \mvec{r}_{\mvec{p}} }
  \Big( \frac{ e^{i \frac{\omega}{c}
      \abs{\mvec{x}-\scrlattb{r}{p}{s}}}}{\abs[\big]{\mvec{x}-\lattb{r}{p}{s}}}
  - \frac{ e^{- i \frac{\omega}{c}
      \abs{\mvec{x}-\scrlattb{r}{p}{s}}}}{\abs[\big]{\mvec{x}-\lattb{r}{p}{s}}}
  \Big) \lattb{U}{}{s} \bigg]  + \\ 
+ \, \bigg[ q^{\scripts{(j)}} e^{i \mvec{k} \cdot
    \mvec{r}_{\mvec{h}} } \Big( \frac{ e^{i \frac{\omega}{c}
      \abs{\mvec{x}-\scrlattb{r}{h}{j}}}}{\abs[\big]{\mvec{x}-\lattb{r}{h}{j}}}
  - \frac{ e^{- i \frac{\omega}{c}
      \abs{\mvec{x}-\scrlattb{r}{h}{j}}}}{\abs[\big]{\mvec{x}-\lattb{r}{h}{j}}}
  \Big) \lattb{U}{}{s} \bigg]\ \, \Bigg\}\ .
\end{multline*}
A classical computation, first carried out by Dirac
\cite{dirac} (see also \cite{marino2}), shows that the second term tends to the radiation
reaction force. Since the first one is continuous at
$\lattb{r}{h}{j}$, its limit must equal the opposite of the radiation
reaction force. Now we observe that this quantity coincides with
\[
\frac{1}{q^{\scripts{(j)}}} e^{i (\mvec{k} \cdot \latt{r}{h} - \omega
  t )} \sum_{s=1}^{n_c} \frac{1}{2} \Big( \hat{\mathcal{D}}_{js} -
\hat{\mathcal{D}}_{sj}^{\dagger} \Big) \cdot \lattb{U}{}{s} .
\]
To see this, factor $e^{i \mvec{k} \cdot \latt{r}{h} } $ out of the
sum, rename the summation index $\mvec{p} - \mvec{h} \mapsto
\mvec{p'}$ and compare the resulting expression with the corresponding
term in \eqref{eq:normalmodes}. At last, it is not difficult to see
that exchanging the indices $j$ and $s$ in the advanced field term
yields the complex conjugate of the retarded one. This term amounts to
one-half the difference between the retarded and the advanced fields
generated by all ions but one, evaluated at the excluded ion site. 
When multiplied by the charge $q^{\scripts{(j)}}$, it becomes equal to the
non-hermitian part of the dipole forces matrix. Therefore, the proof
is complete: the unexpected cancellation appears now better justified
from a theoretical point of view.

If one looks at the book of Born and Huang\cite{bornhuang}, 
one will see that, in
dealing with the secular equation, they take into consideration only
the real part of the equation, which is the one that actually produces
the dispersion relation. Apparently they do not exploit the fact that 
the imaginary part  identically vanishes if the contribution of
the radiation reaction term is taken into account, and just altogether
neglect
the consideration of the imaginary part, as if did not exist. 
A reading of Born's
book \cite{optik} of the year 1933 (see page 431) shows that
the relevance of the classical radiation reaction force was well appreciated by
him. However, he  had to take into account the fact  that it was not easy to
 fit such a force within quantum theory\footnote{In the very words of Born: 
\emph{``Diese ganze klassische Theorie der Strahlungsd\"ampfung ist 
nat\"urlich mit der  heutigen  Quantentheorie des Licht und der
Materie nicht vertr\"aglich.''}}.

\section{Proof of Ewald's formula}

\label{app:ewald}

In this appendix we prove  Ewald's resummation formula, i.e., the formula
which expresses the part of the field due to the  
``far'' dipoles as a rapidly convergent series over the reciprocal lattice. 
Usually, in solid state physics   Ewald's resummation formula is used
in the static limit $\omega\to 0$, i.e.,  in order to resum the Coulomb fields of the far
ions. We report here a proof of the full formula.

So, let us begin  considering the following series
\begin{equation}\label{eq:ewaldmod1}
\Psi(\mvec{x}) \definedas \sum_{\mvec{p}\in \mathbb{Z}^3} e^{ + i
  \mvec{k} \cdot \scrlatt{r}{p} } \frac{e^{i \frac{\omega}{c}
    \abs{\mvec{x}-\latt{r}{p}}}}{\abs[\big]{\mvec{x}-\latt{r}{p}}} \ ,
\end{equation}
which is the series that enters
formula (\ref{eq:fieldpropagation}) of Section~\ref{sec:optics} for the
electric field. Moreover, it enters also  formula 
(\ref{eq:forzadip}) of Section~\ref{sec:matrix} for the dipole matrix
$\hat{\mathcal{D}}_{jl}$, inasmuch as one has
\begin{multline*}
\hat{\mathcal{D}}_{jl}  =
q^{\scripts{(j)}}  \sum_{\mvec{p}\in \mathbb{Z}^3} q^{\scripts{(l)}}
\hat{\mathcal{R}}\left(  \frac {e^{- i \mvec{k} \cdot \scrlatt{r}{p}} \, e^{ i \frac{\omega}{c} \abs{\mvec{x}}}}
        {\abs{\mvec{x}}} \right) \bigg\vert_{\mvec{x}=\scrlattb{r}{p}{j,l}}
         \\
= q^{\scripts{(j)}}    q^{\scripts{(l)}}  \,
  \hat{\mathcal{R}}\Big(\Psi(\mvec{x}+\lattb{b}{}{j,l}) \Bigg) \bigg\vert_{\mvec{x}=0} \,  \ . 
\end{multline*}
So, to prove both equation (\ref{eq:ewalddipolefar})  of Section~\ref{sec:matrix} 
and equation (\ref{eq:splitphi}) of Section~\ref{sec:optics}, one needs to
prove that 
\begin{equation}\label{eq:split}
  \begin{split}
    \Psi( \mvec{x} ) &= 
    \frac{4\pi}{\abs[\big]{V_c}} \sum_{\mvec{m} \in \mathbb{Z}^3} \frac{e^{-
        \frac{1}{4\delta'^2} \big( \abs[\big]{\latt{q}{m} - \mvec{k}}^2 -
        \frac{\omega^2}{c^2} \big) }}{\abs[\big]{\latt{q}{m} - \mvec{k}}^2
      - \frac{\omega^2}{c^2} } e^{i (\mvec{k} - \latt{q}{m} ) \cdot
      \mvec{x} } \\
    &+ \sum_{\mvec{h} \in \mathbb{Z}^3}
    \bigg( e^{ i \mvec{k} \cdot \latt{r}{h}} \frac{2}{\sqrt{\pi}}
    \int_{\delta'}^{+\infty} e^{\big( \frac{\omega^2}{4c^2}\frac{1}{\eta^2} - 
      \abs{\mvec{x}-\latt{r}{h}} \eta^2 \big)} d\eta \bigg)  \ ,
  \end{split}
\end{equation}
with $\delta'$ an arbitrary positive parameter.

To this end, we first reduce the series defining $\Psi(\mvec{x})$ to a
series  over the reciprocal lattice, by using the identities already
introduced in Appendix~\ref{app:WF} 
\begin{gather*}
\frac{e^{\pm i \alpha x}}{x} = 4 \pi \lim_{\epsilon \to 0^+}
\int_{\mathbb{R}^3} d^3 k' \frac{e^{i \mvec{k'} \cdot
    \mvec{x}}}{\abs{\mvec{k'}}^2 - (\alpha^2 \pm i \epsilon)} \, , \\
\frac {(2\pi)^3}{\abs{V_c}} \sum_{\mvec{m} \in \mathbb{Z}^3}\delta(\mvec{x} - \latt{q}{m}) =
\sum_{\mvec{p} \in \mathbb{Z}^3} e^{i
  \scrlatt{r}{p} \cdot \mvec{x}} \, ,
\end{gather*}
$V_c$ being the cell volume, while  $\vect q_{\vect m}$ are  the vectors of
the reciprocal lattice. We recall that,
given a lattice of points $\vect x_{\vect p}$ in a vector space, 
the reciprocal lattice $\vect k_{\vect m}$ is the set of vectors of the dual
space, namely, the vectors such that $\langle\vect x_{\vect p},\vect k_{\vect m}  \rangle$ is an
integer multiple of $2\pi$ (or zero). In $\mathbb{R}^3$, if $\vect a_i$,
$i=1,2,3$, is a basis for the direct lattice, the vectors $\tilde{\vect a}_k
=(2\pi /\abs{V_c})  \big(\vect a_i\wedge \vect a_j\big)$ constitute  a basis for the
reciprocal lattice. Using the mentioned identities one gets
\begin{eqnarray}
\Psi(\mvec{x}) &=&
\sum_{\mvec{p}\in \mathbb{Z}^3}   4 \pi \lim_{\epsilon \to 0^+}
\int_{\mathbb{R}^3} d^3 k' \, \frac{ e^{ i (\mvec{k}-\mvec{k'})
    \cdot \scrlatt{r}{p}} e^{ - i \mvec{k'} \cdot \vect x }}
{\abs{\mvec{k}}^2 - (\omega^2/c^2 +i\epsilon)} \nonumber \\ 
&=& 4 \pi \lim_{\epsilon \to 0^+} \int_{\mathbb{R}^3} d^3 k' 
\frac{ e^{ - i \mvec{k'} \cdot \vect x }}{\abs{\mvec{k}}^2 -
  (\omega^2/c^2 +i\epsilon) } \sum_{\mvec{p}\in \mathbb{Z}^3}  e^{ i (\mvec{k}-\mvec{k'}) \cdot
  \scrlatt{r}{p}} \nonumber \\
&=& \frac{4 \pi}{\abs{V_c}}  \lim_{\epsilon \to 0^+} \int_{\mathbb{R}^3} d^3 k' 
\frac{ e^{ - i \mvec{k'} \cdot \vect x}}{\abs{\mvec{k}}^2 -
  (\omega^2/c^2 \pm i \epsilon)}  
\sum_{\mvec{m} \in \mathbb{Z}^3}\delta(\mvec{k} -\mvec{k'} -
\latt{q}{m})  \nonumber \\
&=& \frac{4 \pi}{\abs[\big]{V_c}} \sum_{\mvec{m} \in \mathbb{Z}^3}
  \frac{e^{i ( \latt{q}{m} - \mvec{k})\cdot \mvec{x} }}
{\abs[\big]{\mvec{k} - \latt{q}{m}}^2 - \omega^2/c^2 } \ .
\end{eqnarray}
For any  $\delta'$, the  series over the reciprocal lattice  
can be conveniently split as follows 
\begin{eqnarray*}
\Psi(\mvec{x}) &=& \Psi_1(\mvec x) +\Psi_2(\mvec x) \definedas \\ 
&~&\frac{4\pi}{\abs[\big]{V_c}} \sum_{\mvec{m} \in \mathbb{Z}^3} \frac{e^{-
    \frac{1}{4\delta'^2} \big( \abs[\big]{\latt{q}{m} - \mvec{k}}^2 -
    \frac{\omega^2}{c^2} \big) }}{\abs[\big]{\latt{q}{m} - \mvec{k}}^2
  - \frac{\omega^2}{c^2} } e^{i (\mvec{k} - \latt{q}{m} ) \cdot
  \mvec{x} }  \\
&+& \frac{4\pi}{\abs[\big]{V_c}} \sum_{\mvec{m} \in \mathbb{Z}^3} \frac{1-e^{-
    \frac{1}{4\delta'^2} \big( \abs[\big]{\latt{q}{m} - \mvec{k}}^2 -
    \frac{\omega^2}{c^2} \big) }}{\abs[\big]{\latt{q}{m} - \mvec{k}}^2
  - \frac{\omega^2}{c^2} } e^{i (\mvec{k} - \latt{q}{m} ) \cdot
  \mvec{x} } \ ,
\end{eqnarray*}
where now the first series converges absolutely, while the second one
can be expressed as an absolutely convergent series by going  back to the
direct lattice.  In fact, using
$$
\frac{1-e^{-
    \frac{1}{4\delta'^2} \big( \abs[\big]{\latt{q}{m} - \mvec{k}}^2 -
    \frac{\omega^2}{c^2} \big) }}{\abs[\big]{\latt{q}{m} - \mvec{k}}^2
  - \frac{\omega^2}{c^2} }
= \int_0^{1/4\delta'^2} \dif \xi \,
e^{-\xi \big( \abs[\big]{\latt{q}{m} - \mvec{k}}^2 -
  \frac{\omega^2}{c^2} \big) } 
$$
together with the identity (see below)
\begin{equation}\label{eq:acci}
\frac{(4 \pi \xi)^{3/2}}{\abs[\big]{V_c}} \sum_{\mvec{m} \in
  \mathbb{Z}^3} 
e^{-  \abs{\latt{q}{m} - \mvec{k}}^2  \xi } \, e^{i \latt{q}{m}  \cdot \mvec{x}} =   
 \sum_{\mvec{h} \in \mathbb{Z}^3} e^{-\frac{1}{4\xi}
   \abs{\mvec{x}-\latt{r}{h}}^2} \, 
e^{ i \mvec{k} \cdot (\mvec{x} - \latt{r}{h})}\ ,
\end{equation}
one obtains
$$
\Psi_2(\mvec x) =
\sum_{\mvec{h} \in \mathbb{Z}^3} \int_0^{1/4\delta'^2} \dif \xi \,
(4 \pi \xi)^{3/2} e^{-\frac{1}{4\xi}
   \abs{\mvec{x}-\latt{r}{h}}^2} \, 
e^{ i \mvec{k} \cdot (\mvec{x} - \latt{r}{h})}
$$
and putting $\eta^2=1/4\xi$ one gets
\begin{equation*}
\Psi_2({\mvec x}) = \sum_{\mvec{h} \in \mathbb{Z}^3}
\bigg( e^{ i \mvec{k} \cdot \latt{r}{h}} \frac{2}{\sqrt{\pi}}
\int_{\delta'}^{+\infty} e^{\big( \frac{\omega^2}{4c^2}\frac{1}{\eta^2} - 
  \abs{\mvec{x}-\latt{r}{h}} \eta^2 \big)} d\eta \bigg)  \ ,
\end{equation*}
which is the relation (\ref{eq:split}).
To complete the proof, there remains  to prove
identity (\ref{eq:acci}). This, however, is in fact rather straightforward, once
one realizes that both sides of it are quasiperiodic functions of $\mvec{x}$. 
One just has to show that the plane-wave coefficients at the 
left-hand side are indeed the Fourier coefficients of the function at 
the right-hand side, that we name $f(\mvec{x})$ for convenience. Hence:
\[
\widetilde{f}_{\mvec{m}}  = \int_{V_c} \frac{d^3 y}{\abs[\big]{V_c}} 
e^{- i \latt{q}{m} \cdot \mvec{y} } f(\mvec{y}) 
 =  \frac{1}{\abs[\big]{V_c}} \sum_{\mvec{h} \in \mathbb{Z}^3}  \int_{V_c}  
e^{- i \latt{q}{m} \cdot \mvec{y} } \, e^{-\frac{1}{4\xi}
  \abs{\mvec{y}-\latt{r}{h}}^2} \,  
e^{ i \mvec{k} \cdot (\mvec{y} - \latt{r}{h})}  \, d^3 y \, .
\]
We now apply the change of variable $\mvec{y'} = \mvec{y}-\latt{r}{h} $, 
so that the integrand does no longer depend on $\mvec{h}$:
\[
\begin{split}
 \widetilde{f}_{\mvec{m}}  & =  \frac{1}{\abs[\big]{V_c}} \sum_{\mvec{h} \in \mathbb{Z}^3}  \int_{V_c + \latt{r}{h} }    e^{-\frac{1}{4\xi} \abs{\mvec{y'}}^2} \,  e^{ i (\mvec{k}- \latt{q}{m}) \cdot \mvec{y'} } \,d^3 y' \\
 & =  \frac{1}{\abs[\big]{V_c}} \int_{\mathbb{R}^3 }   
 e^{-\frac{1}{4\xi} \abs{\mvec{y}}^2} \,  e^{ i (\mvec{k}- \latt{q}{m}) \cdot \mvec{y} }\, d^3 y  \\
& =  \frac{1}{\abs[\big]{V_c}}            \prod_{\mu=1}^3              \int_{\mathbb{R} }    
 e^{-\frac{1}{4\xi} {y}_{\mu}^2} \,  e^{ i (\mvec{k}-
   \latt{q}{m})_{\mu} y_{\mu} } \,  dy_{\mu} \ .
\end{split}
\]
Next, we exploit the integral formula
\[
\int_{\mathbb{R}} e^{- \alpha x^2 + \beta x} dx =
\sqrt{\frac{\pi}{\alpha}} e^{\frac{\beta^2}{4 \alpha}}
\]
to finally get
\[
\widetilde{f}_{\mvec{m}} =  \frac{ ( 4 \pi \xi
  )^{3/2}}{\abs[\big]{V_c}}   e^{ - \frac{1}{4 \xi} \abs{\mvec{k}-
    \latt{q}{m}}^2}\ .
\]
Hence we have
\[
 f(\mvec{x}) =  \sum_{\mvec{m} \in \mathbb{Z}^3} \widetilde{f}_{\mvec{m}} e^{i \latt{q}{m} \cdot \mvec{x } } =  \frac{ ( 4 \pi \xi )^{\frac{3}{2}}}{\abs[\big]{V_c}} \sum_{\mvec{m} \in \mathbb{Z}^3} 
e^{ - \frac{1}{4 \xi} \abs{\mvec{k}- \latt{q}{m}}^2}
e^{i \latt{q}{m} \cdot \mvec{x } } \ ,
\]
and the proof is complete. 

\section{Computable form of the dynamical matrix}
\label{app:dynmatrix}

For the sake of completeness, in this appendix we report the 
computable form of the dynamical matrix $\mathcal{A}(\mvec{k},\omega)$ that 
appears at the right-hand side of the generalized secular equation 
\eqref{eq:normalmodes}, i.e., after Ewald's summation has been performed.

The term $\mathcal{P}$ concerning the phenomenological repulsive 
short-distance interaction (first term at the right-hand side of 
\eqref{eq:normalmodes}) is already in a computable form.

For the Coulomb term $\mathcal{C}$, we have
\begin{align*}
\frac{\hat{\mathcal{C}}_{jj}}{q^{\scripts{(j)}}} & =    - \sum_{i\ne j} q^{\scripts{(i)}}  \bigg\{ \frac{4\pi}{\abs{V_c}}   \sideset{}{'}\sum_{\mvec{m} \in \mathbb{Z}^3} \frac{e^{ - \frac{ \abs{\latt{q}{m}}^2}{4\delta^2} }} { \abs{\latt{q}{m}}^2 } \hat{\mathcal{R}}\big[e^{ i \latt{q}{m} \cdot \mvec{x}}\big]_{\mvec{x}=\lattb{b}{}{j,i}} 
       + \sum_{\mvec{h} \in \mathbb{Z}^3} \hat{\mathcal{R}}
\bigg[ \frac{\erfc \big(\delta  \abs{\mvec{x}}\big)}{\abs{\mvec{x}}} \bigg]_{\mvec{x}=\latt{r}{h} + \lattb{b}{}{j,i}}
\bigg\} \\
    & - q^{\scripts{(j)}}  \bigg\{
\frac{4\pi}{\abs{V_c}}   \sideset{}{'}\sum_{\mvec{m} \in \mathbb{Z}^3}
\frac{e^{ - \frac{ \abs{\latt{q}{m}}^2}{4\delta^2} }}
{ \abs{\latt{q}{m}}^2 } \hat{\mathcal{R}}\big[e^{+ i \latt{q}{m} \cdot \mvec{x}}\big]
_{\mvec{x}=\mvec{0}}  
        + \sideset{}{'}\sum_{\mvec{h} \in \mathbb{Z}^3} \hat{\mathcal{R}}
\bigg[ \frac{\erfc \big(\delta  \abs{\mvec{x}}\big)}{\abs{\mvec{x}}} \bigg]_{\mvec{x}=\latt{r}{h}} 
     +
 \hat{\mathcal{R}}
\bigg[ \frac{\erf \big(\delta  \abs{\mvec{x}}\big)}{\abs{\mvec{x}}} \bigg]_{\mvec{x}=\mvec{0}}
\bigg\} \ ; 
\end{align*} 
for all $ j=1,\dots,n$, and
\[
\hat{\mathcal{C}}_{jl} =0 \qquad \text{if } j \ne l \ .
\]
The off-diagonal blocks vanish because the Coulomb term describes 
the force exerted on an ion by all the other charges, 
supposed fixed at their equilibrium positions: no 
\emph{coupling} between two different ions occurs. 

Finally, for the dipole term $\mathcal{D}$, we have
\begin{align*}
\frac{\hat{\mathcal{D}}_{jj}}{ \big(q^{\scripts{(j)}}\big)^2} &=  
\frac{4\pi}{\abs{V_c}}  \sum_{\mvec{m} \in \mathbb{Z}^3}
\frac{e^{ - \frac{ 1}{4\delta'^2} \big( \abs{\latt{q}{m}-\mvec{k}}^2 - \frac{\omega^2}{c^2} \big) }}
{ \abs{\latt{q}{m}-\mvec{k}}^2 - \frac{\omega^2}{c^2}  } \hat{\mathcal{R}}\big[e^{ i (\latt{q}{m}-\mvec{k}) \cdot \mvec{x}}\big]_{\mvec{x}=\mvec{0}}  \\
     & +   \sideset{}{'}\sum_{\mvec{h} \in \mathbb{Z}^3} e^{ - i \mvec{k} \cdot \latt{r}{h}} 
 \frac{2}{\sqrt{\pi}} \int_{\delta'}^{+\infty} e^{\frac{\omega^2}{4c^2}\frac{1}{\eta^2}} \hat{\mathcal{R}} \big[ e^{- \abs{\mvec{x}}^2 \eta^2 } \big]_{\mvec{x}=\latt{r}{h}}  d\eta  \\
   &  +
 \hat{\mathcal{R}}
\bigg[ 
\frac{1}{\abs{\mvec{x}}} 
 \frac{2}{\sqrt{\pi}} \int_{\delta'}^{\infty} e^{\frac{\omega^2}{4c^2}\frac{1}{\eta^2}} e^{- \abs{\mvec{x}}^2 \eta^2 } \, d\eta - e^{ i \frac{\omega}{c} \abs{\mvec{x}}} 
 \bigg]_{\mvec{x}=\mvec{0}}
\end{align*} 
for all $j=1,\dots,n$, and
\begin{align*}
\frac{\hat{\mathcal{D}}_{jl}}{q^{\scripts{(j)}} q^{\scripts{(l)}}}  &=   
\frac{4\pi}{\abs{V_c}}  \sum_{\mvec{m} \in \mathbb{Z}^3}
\frac{e^{ - \frac{ 1}{4\delta'^2} \big( \abs{\latt{q}{m}-\mvec{k}}^2 - \frac{\omega^2}{c^2} \big) }}
{ \abs{\latt{q}{m}-\mvec{k}}^2 - \frac{\omega^2}{c^2}  } \hat{\mathcal{R}}\Big[e^{ i (\latt{q}{m}-\mvec{k}) \cdot \mvec{x}}\Big]_{\mvec{x}=\lattb{b}{}{j,l}} \\
      & +  \sum_{\mvec{h} \in \mathbb{Z}^3} e^{ - i \mvec{k} \cdot \latt{r}{h}} 
 \frac{2}{\sqrt{\pi}} \int_{\delta'}^{\infty} e^{\frac{\omega^2}{4c^2 \eta^2}} \hat{\mathcal{R}} \Big[ e^{- \abs{\mvec{x}}^2 \eta^2 } \Big]_{\mvec{x}=\lattb{r}{h}{j,l}}  d\eta 
\end{align*} 
$ \text{for all } j,l=1,\dots,n, \; j \ne l$.

It can be seen that the total electric term (Coulomb plus dipole) 
takes the form of a standard ``mechanical'' dynamical matrix in 
the \emph{instantaneous} limit, the potential being
\[
\phi_{\scripts{(j,l)}} ( r ) = \frac{q^{\scripts{(j)}}
  q^{\scripts{(l)}} }{r}\  .
\]
As already pointed out, this limit is formally obtained by putting 
$\omega=0$ in the matrix elements of $\mathcal{D}$. 
Thereby, the electric force matrix $\mathcal{C}+\mathcal{D}$ 
(actually its hermitian part) reduces to a ``mechanical'' dynamical
matrix, analogous to $\mathcal{P}$, the only difference being related to 
the long range of the interaction. However, the purpose of this work
is to investigate precisely those situations where the instantaneous 
limit loses validity.


\end{document}